# Doping and Temperature Dependence of Raman Scattering of NdFeAsO$_{1-x}$F$_x$ (x=0.0, 0.1, 0.2) Superconductor

L. Zhang [1], T. Fujita [1], F. Chen [2], D. L. Feng [2], S. Maekawa [3,4] and M. W. Chen [1,3,*]

[1] World Premier International (WPI) Research Center, Advanced Institute for Materials Research, Tohoku University, Sendai 980-8577, Japan

[2] Department of Physics, Fudan University, Shanghai 200433, China

[3] Institute for Materials Research, Tohoku University, Sendai 980-8577, Japan

[4] CREST Japan Science Technology Agency, Tokyo 102-0075, Japan

**Raman spectra of polycrystalline NdFeAsO$_{1-x}$F$_x$ (x=0.0, 0.1, 0.2) compound have been systematically investigated as functions of temperature and fluorine concentration. Scanning electron microscopic and Raman microscopic characterization demonstrates that the polycrystalline samples mainly contain two phases, *i.e.* superconductor NdFeAsO$_{1-x}$F$_x$ compound and a MnP-type FeAs phase, with dissimilar characteristic Raman bands. It was found that fluorine doping leads to structure disorder in the insulator Nd-O layers and high temperature coefficient of Fe-As vibrational mode.**

*Corresponding author: Mingwei Chen,　E-mail: mwchen@imr.tohoku.ac.jp

The recent discovery of $RFeAsO_{1-x}F_x$ (R: rare earth elements) superconductors has stimulated the tremendous research interests since they are the first example of a noncuprate transition metal compound with unconventional superconductivity and high transition temperatures.[1-8] According to traditional BCS theory, electrons form Cooper pairs through an interaction mediated by vibrations of crystal lattices. Thus, although superconductivity is an electronic phenomenon, it is more or less associated with the vibrations (phonons) of crystal lattices in which the electrons travel. Although the underlying micromechanism of high Tc superconductivity of the new materials appears much more complex than that described by the traditional phonons medicated Cooper pairs, Raman spectroscopy can still offer unique insights on the mechanism of superconductivity by detecting phonons or other excitations in the new superconductor compounds. Moreover, Raman spectroscopy is also a powerful tool in identifying various phases in polycrystalline inorganic materials and in studying underlying atomic and electronic structure changes caused by minor doping. In this study, we systematically investigated the microstructure and phonon modes of polycrystalline $NdFeAsO_{1-x}F_x$ (x=0.0, 0.1, 0.2) compound and found that fluorine doping results in visible variations in atomic structure and phonon excitations of the Fe-based superconductors.

Polycrystalline samples with nominal compositions of $NdFeAsO_{1-x}F_x$ (x=0.0, 0.1, 0.2) were synthesized by a standard two-step solid state reaction using NdAs, $NdF_3$, FeAs, Fe and $Fe_2O_3$ (purity >99.9%) as green powders. The samples with the stoichiometric ratio of $NdFeAsO_{1-x}F_x$ (14:1:1:6:4 for x=0.2; 29:1:1:11:18 for x=0.1) were sintered at 1150 °C for 48 hours. Resistivity measurements demonstrate that two fluorine doped samples, $NdFeAsO_{0.9}F_{0.1}$ and $NdFeAsO_{0.8}F_{0.2}$, are superconductors with Tc of about 33 and 51K, respectively. The sample surfaces were carefully polished for scanning electron microscopic (SEM) characterization and for Raman scattering measurements. Crystalline phases in the sintered samples were characterized by a X-ray diffraction (XRD) spectrometer with Cu K$\alpha$ radiation ($\lambda = 1.5406$ Å). Microstructure of the polycrystalline samples was inspected by SEM (JOEL JSM-7000F) and an optical microscope. Raman studies were performed by using a Raman microscope (Renishaw InVia RM 1000) with an incident wavelength of 632.8 nm. The power of the laser is set at a low value of $5 \times 10^4$ W/cm$^2$ to avoid possible damage by laser irradiation. Low-temperature Raman study was conducted using a heating/cooling stage where liquid nitrogen (LN) was used as a cryogenic source and a resistance heater for heating. [9] The sample temperatures were measured using a Pt-based thermal couple with an accuracy of ±0.1 K.

XRD patterns of three $NdFeAsO_{1-x}F_x$ samples are illustrated in Figure 1. Most of the diffraction peaks in the XRD spectra can be assigned to a tetragonal ZrCuSiAs-type (P4/nmm) phase, except for some weak peaks from secondary impurity phases. [1,7] Slight peak shift caused by fluorine doping, corresponding to the lattice constant changes, can be detected. Moreover, the XRD spectra indicate that the fluorine doped samples appear to contain more secondary phases. Back-scattering SEM micrographs reveal that there are mainly two phases with difference contrast in the polycrystalline samples. Fig. 2 (a) shows an example of the microstructure of the $NdFeAsO_{0.9}F_{0.1}$ sample in which chemical compositions of individual phases are measured by SEM energy dispersive X-ray spectroscopy (EDS). The white phase contains all the nominal components of Nd, Fe, As,

F and O and thus is the NdFeAsOF superconductor. In contrast, the dark phase is only composed of Fe and As. Combining with XRD and electron diffraction (not shown here), the secondary phase is determined to be a FeAs phase with a MnP-type crystal structure.

With the known microstructure of the polycrystalline samples, Raman spectra of the two phases were measured from the polished samples using the Raman microscope. The laser beam size used in this study is about 2 µm, much smaller than grain sizes of the NdFeAsO$_{1-x}$F$_x$ superconductor and the FeAs phase. Thus, single phase spectra can be acquired in our study. The characteristic Raman spectra of the superconductor are comprised of three major bands at 487cm$^{-1}$, 338 cm$^{-1}$ and 262cm$^{-1}$(Fig 2C, line a) as well as a few of minor peaks. According to the literature, the 487cm$^{-1}$ band comes from Nd-O stretching mode,[10-12] 338 cm$^{-1}$ from oxygen mode along axis direction,[10, 12] and 262cm$^{-1}$ from vibrational mode of Fe-As bond.[13-15] These bands and their assignments match well with the known atomic structure of the NdFeAsO$_{1-x}$F$_x$ superconductor. For the FeAs phase, there are two characteristic vibrational modes (Fig 2C, line b) at 242cm$^{-1}$ and 271cm$^{-1}$. Both of them are mixed stretching and librational vibrations of Fe-As bond.[16] It is worth emphasizing that the Raman scattering of the FeAs phase is much stronger than that of the superconductor phase, which easily causes confusion when one does not have the pre-knowledge of the microstructure of the polycrystalline materials.

It has been known that fluorine doping is the key leading to the metal-to-superconductor transition in the ReFeAsO compounds at high temperatures. Thus, detecting the doping effects in atomic structure and phonon excitations may offer useful information on understanding the superconductivity of the new Fe-based materials. Fig. 2 shows the Raman spectra of NdFeAsO$_{1-x}$F$_x$ (x=0.0, 0.1, 0.2) and the band positions and assignments of optical phonons are listed in Table 1. Visible band shift and intensity changes of the three characteristic peaks of the superconductor phase can be detected with the change of fluorine/oxygen ratios. The most notable variation caused by fluorine doping is the dramatic loss in the intensity of the Nd-O stretching mode at ~485cm$^{-1}$. Meanwhile, this peak becomes broad and asymmetrical. Thus, it can be concluded that the doped fluorine atoms substitute oxygen in the Nd-O layers and lead to structure disorder in the insulated oxide layers. The broad and asymmetrical Nd-O band also indicates that the distribution of fluorine in the NdO layers is most likely random. Unlike the Nd-O mode, the intensity and shape of the Raman bands related to Fe-As layers do not show visible change. However, frequency shifts of the three characteristic bands related to both Fe-As and Nd-O bonds occur with fluorine doping. Careful measurements of the band positions demonstrate that the Nd-O stretching mode at 485cm$^{-1}$ shifts to 490cm$^{-1}$ with ~5 at. % fluorine doping, suggesting that the bond distance of Nd-O decreases and the binding energy increases with the substitution of oxygen by fluorine. Although the structure changes mainly take place in the oxide layers, fluorine doping also causes discernible structure changes in the FeAs layers as evidenced by the slight downshift of the vibrational mode of Fe-As bond from 262.7 to 261.1 cm$^{-1}$ with ~5 at. % fluorine doping, indicating slight increase in the bond distance and decrease in binding energy. Moreover, a weak peak at ～368cm$^{-1}$ gradually emerges with fluorine doping. But, the origin of this mode is not clear. Probably it is related to the vibration of Nd-O bond or Nd-F bond.[10]

Since superconducting transition occurs at low temperatures, temperature dependence of phonon vibrations of superconductor compounds may offer valuable clues on the mechanism of superconductivity. The Raman spectra of the $NdFeAsO_{1-x}F_x$ compound at temperatures ranging from 90 to 300 K are shown in Fig. 4. In general, as temperature decreases, the structure of a compound becomes more compact with the decrease of bond distance, leading to Raman bands shift to high frequencies. The temperature-induced band shift of the $NdFeAsO_{1-x}F_x$ compound exhibits the similar behavior. However, the temperature dependence of the Raman scattering of the $NdFeAsO_{1-x}F_x$ compound is anisotropic and the three characteristic modes have difference temperature coefficients. As shown in Fig 5, in the parent NdFeAsO compound the temperature coefficient of the Fe-As mode at ~262cm$^{-1}$ is about $(2.12\pm0.07)\times10^{-2}$, which is about two times larger than that of Nd-O mode at ~487cm$^{-1}$ ($(1.10\pm0.11)\times10^{-2}$). Thus, the conducting Fe-As layers are more sensitive to temperature changes than the insulator oxide layers, which is in agreement with the fact that metals generally have higher temperature coefficients than insulator ceramics. Additionally, fluorine doping enlarges the difference in the temperature coefficients of the three characteristic modes of the superconductor compound. As we discussed before (Fig. 3), the changes in the structures and phonon vibrations of $NdFeAsO_{1-x}F_x$ with fluorine doping are mainly associated with the Nd-O insulator layers because of the disorder caused by the substitution of oxygen by fluorine. Interestingly, the temperature coefficient of Fe-As bond is much more susceptible to the fluorine doping than that of Nd-O bond. The temperature coefficient of the Fe-As mode increases from $(2.12\pm0.07)\times10^{-2}$ to $(3.24\pm0.26)\times10^{-2}$ with ~5 at.% fluorine doping whereas the temperature coefficient of Nd-O bond only slightly changes from $(1.10\pm0.11)$

×10$^{-2}$ to (1.45±0.15) ×10$^{-2}$. The high temperature dependence of the conducting Fe-As layers is probably related to electron-phonon interaction and electron doping through the replacement of O$^{2-}$ by F$^{-}$ enhances this interaction and leads to the Fe-As layers more metal-like.

In summary, we systematically investigated the Raman scattering of the polycrystalline NdFeAsO$_{1-x}$F$_x$ materials. With the assistance of SEM microstructure characterization, the characteristic Raman bands of the NdFeAsO$_{1-x}$F$_x$ superconductor were clarified. The fluorine doping was found to cause the structure disorder of the insulator Nd-O layers and thereby dramatic loss in the intensity of the Nd-O stretching mode. Additionally, the fluorine doping also leads to the detectable downshift and increased temperature coefficient of the Fe-As vibrational mode, demonstrating that the electron doping by F$^{-}$ substituting O$^{2-}$ results in the weakening of Fe-As bonding and the more metal-like behavior of the conducting Fe-As layers.

[1] Y. Kamihara, T. Watanabe, M. Hirano, and H. Hosono, J. Am. Chem. Soc. **130**, 3296 (2008).

[2] H. Takahashi, K. Igawa, K. Arii, Y. Kamihara, M. Hirano, and H. Hosono, Nature **453**, 376 (2008).

[3] C. Cruz, Q. Huang, J. W. Lynn, J. Li, W. Ratcliff, J. L. Zarestky, H.A. Mook, G. F. Chen, J. L. Luo, N. L. Wang, and P. Dai, Nature **453**, 899(2008).

[4] G. F. Chen, Z. Li, D. Wu, G. Li, W. Z. Hu, J. Dong, P. Zheng, J. L. Luo, and N. L. Wang, arxiv:0803.3790v2.

[5] X. H. Chen, T. Wu, G. Wu, R. H. Liu, H. Chen, and D. F. Fang, Nature, **453**, 761(2008).

[6] F. Hunte, J. Jaroszynski, A. Gurevich, D. C. Larbalestier, R. Jin, A. S. Sefat, M. A. McGuire, B. C. Sales, D. K. Christen, and D. Mandrus, Nature **453**, 903(2008).

[7] J. Ying, P. Cheng, L. Fang, H. Luo, H. Yang, C. Ren, L. Shan, C. Gu, and H. Wen, Appl. Phys. Lett. **93**, 032503 (2008).

[8] H. J. Grafe, D. Paar, G. Lang, N. J. Curro, G. Behr, J. Werner, J. Hamann-Borrero, C. Hess, N. Leps, R. Klingeler, and B. Büchner, Phys. Rev. Lett. **101**, 047003(2008).

[9] X. Q. Yan, W. J. Li, T. Goto, and M. W. Chen, Appl. Phys. Lett. **88**, 131905(2006).

[10] V. G. Hadjiev, M. N. Iliev, K. Sasmal, Y. Y. Sun, and C. W. Chu, Phys. Rev. B **77**, 220505(R)(2008).

[11] S. Mozaffari, and M. Akhavan, Physica C **468**, 985(2008).

[12] Y. Morioka, A. Tokiwa, M. Kikuchi, and Y. Syono, Solid State Communications **67**, 267(1988).

[13] S. C. Zhao, D. Hou, Y. Wu, T. L. Xia, A. M. Zhang, G. F. Chen, J. L. Luo, N. L. Wang,

J. H. Wei, Z. Y. Lu, and Q. M. Zhang, arXiv: 0806.0885v1.

[14] J. Dong, H. J. Zhang, G. Xu, Z. Li, G. Li, W.Z. Hu, D. Wu, G. F.Chen, X. Dai, J. L. Luo, Z. Fang, and N.L. Wang, arXiv: 0803.3426v1.

[15] D. J. Singh, and M. H. Du, arXiv: 0803.0429v1.

[16] H. D. Lutz, and B. Müller, Phys. Chem. Minerals **18**, 265(1991).

**Table1. Characteristic Raman bands of NdFeAsO$_{1-x}$F$_x$ compounds.**

| Atom | Vibration | Experimental Position (cm$^{-1}$) | | |
|---|---|---|---|---|
| | | NdFeAsO | NdFeAsO$_{0.9}$F$_{0.1}$ | NdFeAsO$_{0.8}$F$_{0.2}$ |
| Fe, As | mixed mode of Fe and As in *c* direction[13-15] | 253 | 252.5 | 252.1 |
| Fe, As | mixed mode of Fe and As in *a* or *b* direction[13-15] | 262.7 | 262.3 | 261.1 |
| O | oxygen mode along axis direction[10,12] | 338.1 | 337.8 | 336.7 |
| O or F | Nd-O bond [10] or Nd-F bond | 368 | 368 | 364 |
| Nd, O | Nd-O stretching mode [10-12] | 485 | 487 | 490 |

**Note:** *a* and *b* directions are along FeAs layers (parallel to NdO layers) and *c* is the axis direction perpendicular to the *a-b* plane.

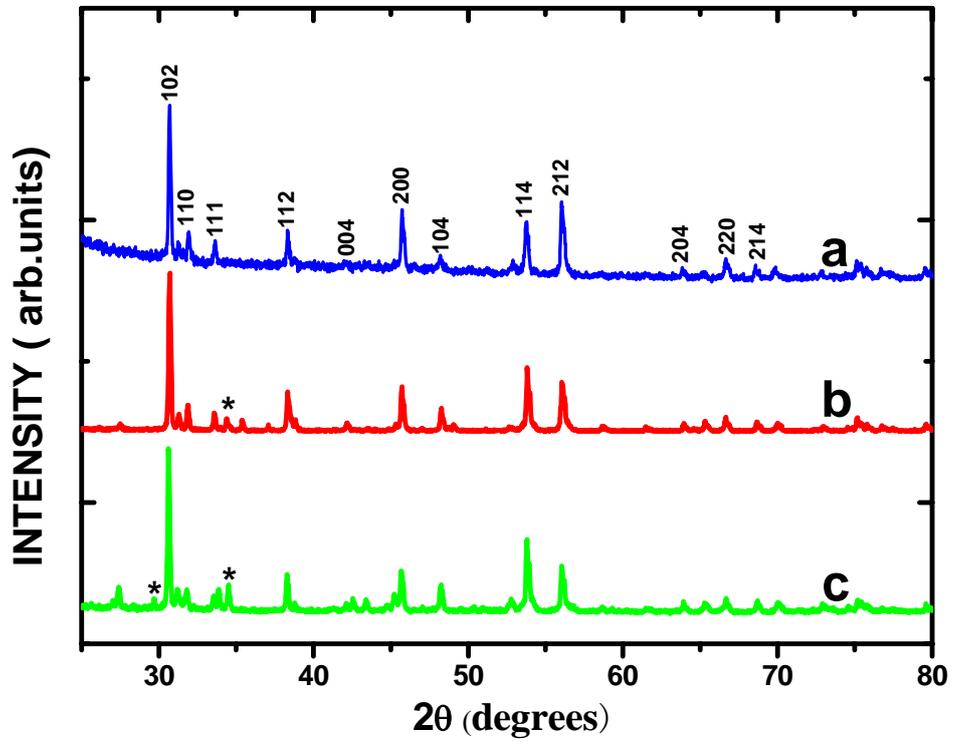

**FIG.1.** XRD patterns of polycrystalline NdFeAsO$_{1-x}$F$_x$ with (a) x=0; (b) x=0.1; and (c) x=0.2. The stars denote the peaks from the secondary FeAs phase.

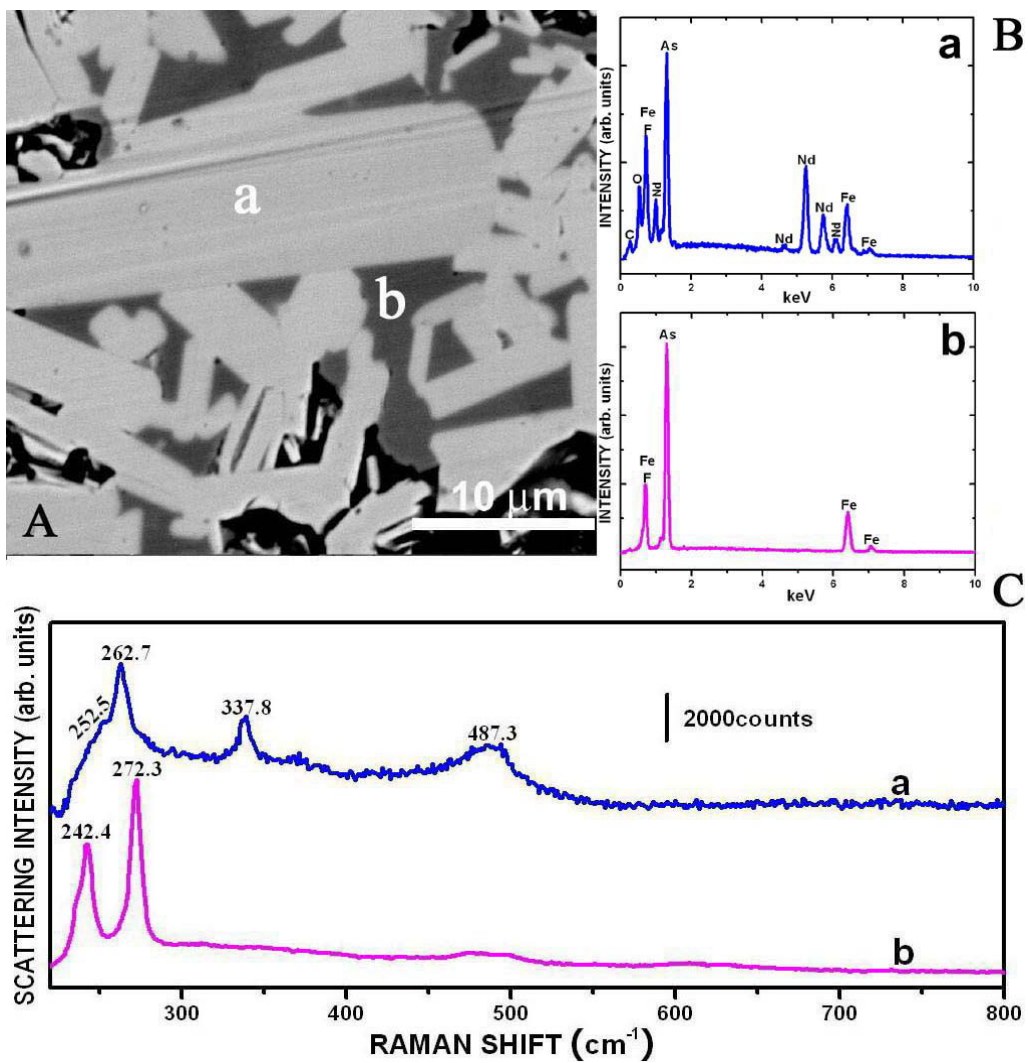

**FIG.2.** (A) Back scattering SEM micrograph of NdFeAsO$_{0.9}$F$_{0.1;}$ (B) and (C) EDX and Raman spectra corresponding to (a) superconducting NdFeAsO$_{0.9}$F$_{0.1}$ compound and (b) the FeAs phase.

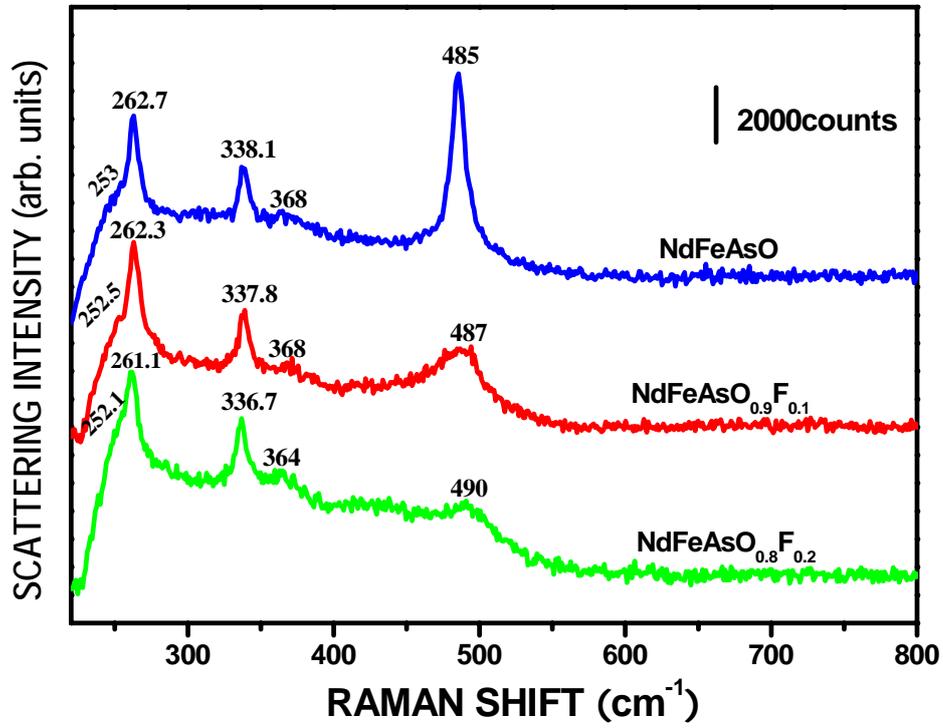

**FIG.3.** Micro-Raman spectra of NdFeAsO$_{1-x}$F$_x$ compounds.

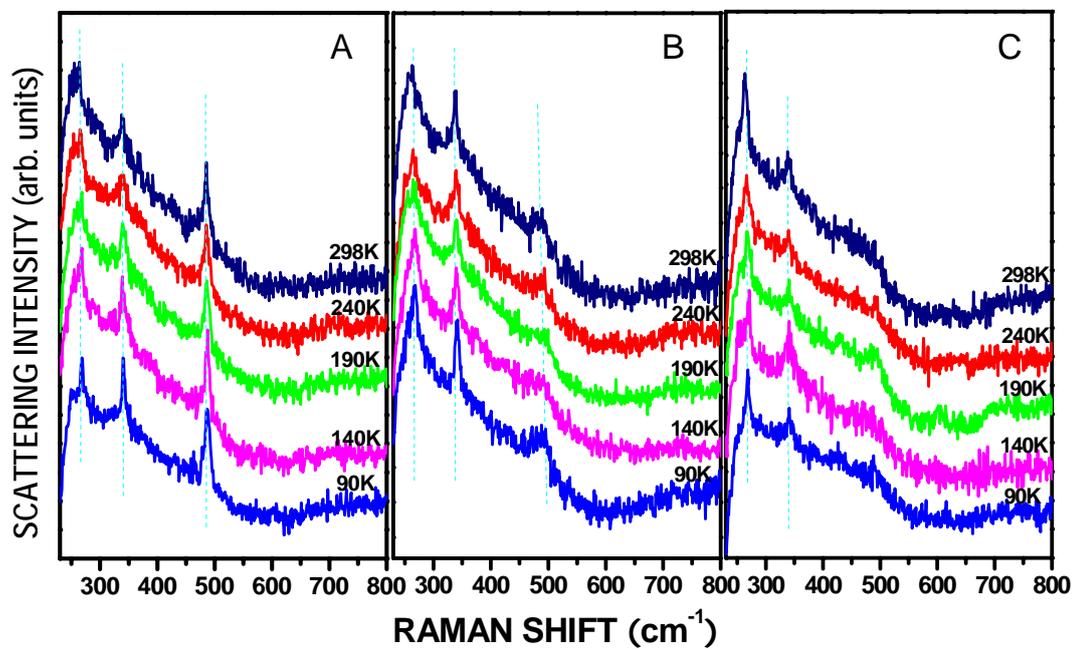

**FIG.4**. Raman spectra of NdFeAsO$_{1-x}$F$_x$ at the temperature ranging from 298K to 90K. (A) NdFeAsO; (B) NdFeAsO$_{0.9}$F$_{0.1}$; and (C) NdFeAsO$_{0.8}$F$_{0.2}$.

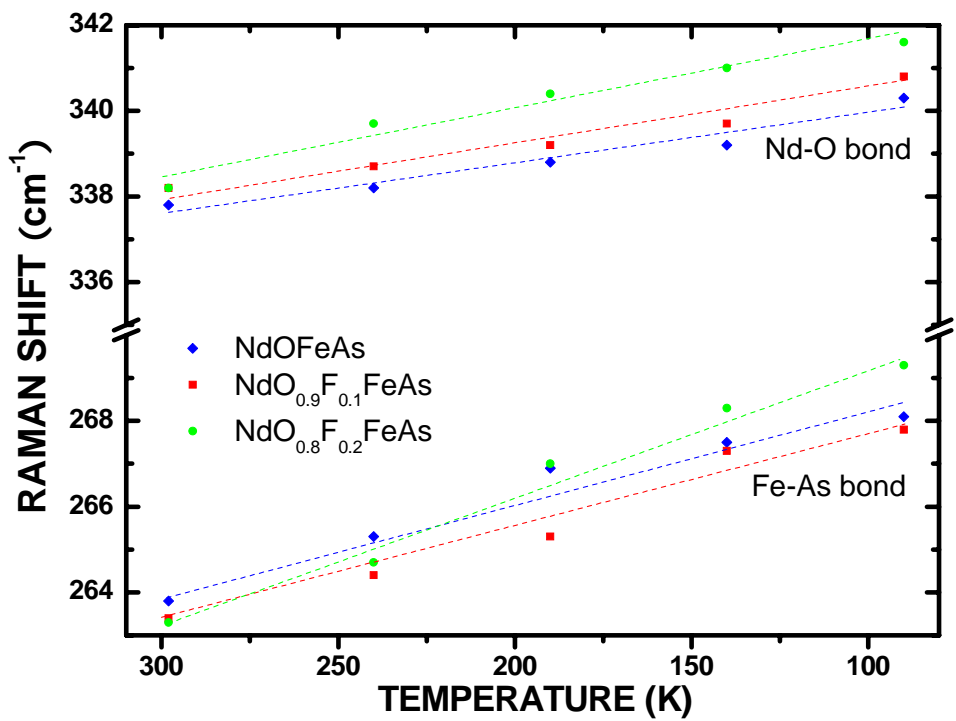

**FIG.5.** Temperature dependence of Raman band shifts of the superconductor phase in the polycrystalline $NdFeAsO_{1-x}F_x$ compounds.